\documentclass{article}

\usepackage{arxiv}

\usepackage[utf8]{inputenc} % allow utf-8 input
\usepackage[T1]{fontenc}    % use 8-bit T1 fonts
\usepackage{hyperref}       % hyperlinks
\usepackage{url}            % simple URL typesetting
\usepackage{booktabs}       % professional-quality tables
\usepackage{amsfonts}       % blackboard math symbols
\usepackage{nicefrac}       % compact symbols for 1/2, etc.
\usepackage{microtype}      % microtypography
\usepackage{graphicx}
\usepackage{epstopdf, epsfig}
\usepackage{amsmath}
\usepackage{graphicx}
\usepackage{graphics}
\usepackage{subfig}

\title{Accelerating Deep Reinforcement
Learning strategies of Flow Control through a multi-environment approach}
\author{
  Jean Rabault\\
  Department of Mathematics\\
  University of Oslo\\
  \texttt{jean.rblt@gmail.com} \\
   \And
  Alexander Kuhnle \\
  University of Cambridge\\
  \texttt{alexkuhnle@t-online.de} \\
}

\begin{document}
\maketitle

\begin{abstract}
Deep Reinforcement Learning (DRL) has recently been proposed as
    a methodology to discover complex Active
    Flow Control (AFC) strategies [Rabault, J., Kuchta, M., Jensen, A.,
    Réglade, U., \& Cerardi, N. (2019): ``Artificial neural networks trained
    through deep reinforcement learning discover control strategies for active
    flow control'', Journal of Fluid Mechanics, 865, 281-302]. However, while
    promising results were obtained on a simple 2D benchmark flow at a moderate
    Reynolds number, considerable speedups will be required to investigate more
    challenging flow configurations. In the case of DRL trained with
    Computational Fluid Dynamics (CFD) data, it was found that the CFD part,
    rather than training the Artificial Neural Network, was the limiting
    factor for speed of execution. Therefore, speedups should be obtained
    through a combination of two approaches. The first one, which is well
    documented in the literature, is to parallelize the
    numerical simulation itself. The second one is to adapt the DRL algorithm
    for parallelization. Here, a simple strategy is to use several independent
    simulations running in parallel to collect experiences faster. In the present work,
    we discuss this solution for parallelization. We illustrate that perfect speedups
    can be obtained up to the batch size of the DRL agent, and slightly suboptimal
    scaling still takes place for an even larger number of simulations. This is,
    therefore, an important step towards enabling the study of more sophisticated
    Fluid Mechanics problems through DRL. 
\end{abstract}

\section{Introduction}

Active Flow Control (AFC) is a problem of considerable theoretical and practical interest, with many applications
including drag reduction on vehicles and airplanes \citep{pastoor2008feedback, YOU20081349, PhysRevFluids4034604,
li2019optimization}, or optimization of the combustion processes taking place in engines \citep{Wu2018}. Unfortunately,
the difficulty of finding efficient strategies for performing AFC is well-known \citep{brunton2015closed, duriez2017machine}. It
arises from the combination of non-linearity, time-dependence, and high dimensionality inherent to the Navier-Stokes
equations.

The importance of AFC is apparent from the large body of literature that
discusses both theoretical, computational, and experimental aspects of the
problem. In particular, active flow control has been discussed in simulations
using both Reduced Order Models and harmonic forcing
\citep{doi:10.1063/1.2033624}, direct simulations coupled with the adjoint
method \citep{doi:10.1063/1.4928896} or linearized models
\citep{doi:10.1063/1.1359420}, and mode tracking methods
\citep{doi:10.1063/1.5085474}. Realistic actuation mechanisms, such as plasma
actuators \citep{doi:10.1063/1.4935357}, suction mechanisms
\citep{doi:10.1063/1.4947246}, transverse motion \citep{doi:10.1063/1.1582182},
periodic oscillations \citep{doi:10.1063/1.3560379}, oscillating foils
\citep{doi:10.1063/1.4802042}, air jets \citep{doi:10.1063/1.5092851}, or
Lorentz forces in conductive media \citep{doi:10.1063/1.1647142}, are also
discussed in details, as well as limitations imposed by real-wold systems
\citep{doi:10.1063/1.4804390}. Similarly, a lot of experimental work has been
performed, with the goal of controlling either the cavitation instability
\citep{doi:10.1063/1.5099089}, the vortex flow behind a conical forebody
\citep{doi:10.1063/1.5005514}, or the flow separation over a circular cylinder
\citep{doi:10.1063/1.3237151, doi:10.1063/1.3194307}. Finally, while most of
the literature has focused on complex, closed-loop control methods, some
open-loop methods have also been discussed, both in simulations
\citep{doi:10.1063/1.3425625} and in experiments \citep{doi:10.1063/1.5082945}.

However, it is still challenging to apply active flow control to situations of
industrial interest. By reviewing the literature presented in the previous
paragraph one can observe that, while there are good candidates for both
sensors and actuators to be used in such systems, finding algorithms for using
those measurements and actuation possibilities in an efficient way to perform
AFC is challenging. In addition, challenges such as disturbances inherent to the
real world, imperfections in the building of the sensors or actuators, and
adaptivity to changing external conditions may also be part of the problem.
Therefore, the main limitation to AFC is currently the lack of robust,
efficient algorithms that can leverage the physical devices available for
performing such control. The difficulty of finding efficient algorithms for
control and interaction with real world situations or complex systems is also
present in other fields of research such as speech recognition
\citep{jelinek1997statistical}, image analysis \citep{TARR19981}, or playing
complex games such as the game of Go or poker. In those domains, great
progress has recently been obtained through the use of data-driven methods
and machine learning, and problems that had been unattainable for several
decades have been practically solved in recent years \citep{graves2013speech,
krizhevsky2012imagenet, silver2017mastering, brown2019superhuman}.

Since data-driven and learning-based methods are suitable to be used
on non-linear, high dimensional, complex problems, they are, therefore,
also promising for performing AFC \citep{duriez2017machine}. More specifically,
such promising methods include Genetic Programming (GP)
\citep{gautier2015closed, duriez2017machine}, and Deep Reinforcement Learning
(DRL) \citep{Verma201800923, rabault2018deep,
rabault_kuchta_jensen_reglade_cerardi_2019}. Those two methods are among the
most successful data-driven approaches at performing nonlinear control
tasks in the recent years, and are regarded as possible solutions to the
challenges faced by AFC \citep{duriez2017machine}. In
addition, one of their key properties is their ability to naturally scale, in a
parallel fashion, to large amounts of data and / or computational power, which
enables very efficient training. This is a well known fact for GP, due to
its inherently parallel nature. However, it is maybe slightly less known that
the same is true also of DRL, if a suitable implementation is used. In the
present work, we illustrate how parallelization of DRL can be performed by
extending the work presented in
\citet{rabault_kuchta_jensen_reglade_cerardi_2019}, and we provide a flexible
implementation that will be available open-source. Therefore,
we provide both implementations and guidelines that may be used by future work
as a basis for further investigation of the application of DRL to AFC in 
complex scenarios.

In the following, we first provide a short reminder of the Computational Fluid
Dynamics (CFD) simulation used, and of the main principles behind DRL. Then, we
explain how the data collection part of the algorithm can be parallelized in a
general way. Finally, we present the scaling results obtained and we discuss
their significance for the use of DRL in more challenging AFC situations

\section{Methodology}

In this section, we summarize the methodology for the CFD simulation (subsection \ref{CFD}) and the DRL algorithm
(subsection \ref{PPO}). Those are identical to what was already used in
\citet{rabault_kuchta_jensen_reglade_cerardi_2019}. In addition, we present the method used for
parallelizing the DRL (subsection \ref{parallel}), which is the new contribution of this work.

\subsection{Simulation environment \label{CFD}}

\begin{figure*}[ht]
\begin{center}
\includegraphics[width=.99\textwidth]{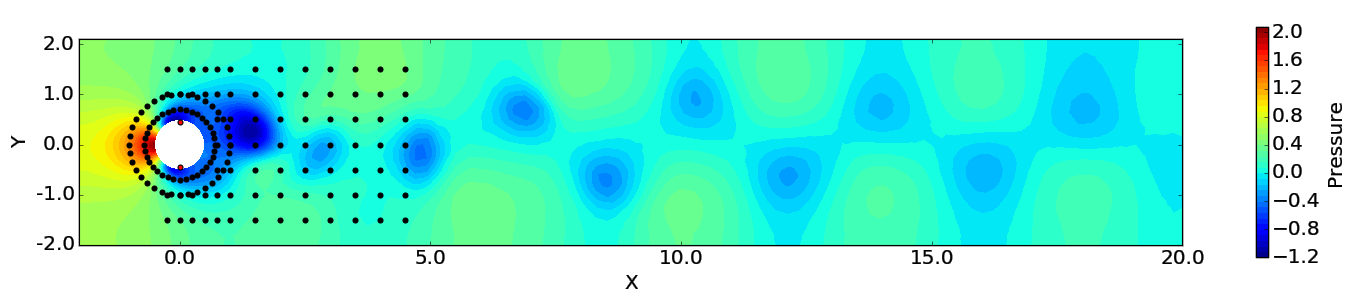}
\caption{\label{fig:general_config} Unsteady non-dimensional pressure wake behind the cylinder after flow
initialization without active control. The location of the pressure probes is indicated by the black dots. The location
of the control jets is indicated by the red dots. This illustrates the configuration used to perform learning.
}
\end{center}
\end{figure*}

The CFD simulation is identical to the one presented in \citet{rabault_kuchta_jensen_reglade_cerardi_2019}, and the
reader interested in more details should refer to this work
and the open-source code implementation: \url{https://github.com/jerabaul29/Cylinder2DFlowControlDRL}. To summarize, the CFD case chosen is a simple 2D simulation of the
non-dimensionalized flow around a cylinder as described by the incompressible Navier-Stokes equations at $Re=100$. The
configuration chosen is a copy of a classical benchmark used for the validation of numerical codes \citep{Schafer1996}.
In addition, two small jets of angular width $10^{\circ}$ are set on the sides of the cylinder and inject fluid in the
direction normal to the cylinder surface, following the control setup by the Artificial Neural Network (ANN). This
implies that the control relies on influencing the separation of the {K}{\'a}rm{\'a}n vortices, rather than direct
injection of momentum as could be obtained through propulsion. The jets are controlled through their mass flow rates,
respectively $Q_1$ and $Q_2$. We choose to use synthetic jets, meaning that no net mass flow rate is injected in the
flow, which translates to the constraint $Q_1 + Q_2 = 0$. The general configuration of the simulation is presented in
Fig. \ref{fig:general_config}.

The governing Navier-Stokes equations are solved in a segregated manner \citep{valen2012comparison}. More precisely, the
Incremental Pressure Correction Scheme (IPCS method, \citet{GODA197976}) with an explicit treatment of the non-linear term
is used. Spatial discretization then relies on the finite element method implemented within the FEniCS framework
\citep{logg2012automated}.

The aim of the control strategy is guided by the reward function fed to the DRL during training. In the present work, we
want to minimize the drag $D$ through a reduction in the strength of the vortex shedding, in a way analogous to what is
obtained by boat tailing \citep{rabault_kuchta_jensen_reglade_cerardi_2019}. This means, in practice, that we are looking
for strategies that use the jets as a way to control the vortex shedding process, rather than a way to perform propulsion or
converting some of the drag into lift through curbing of the wake. For this, we define the reward
function $r$ from both the drag $D$ and the lift $L$, following:

\begin{equation} r = \langle D \rangle_{S} - |\langle L\rangle_{S}|, \label{reward_function} \end{equation}

\noindent where $\langle\bullet\rangle_{S}$ indicates the mean over an action step of the ANN (see section \ref{PPO}).
In Eq. (\ref{reward_function}), the term related to drag penalizes large drag values (which are negative due to conventions
used in the code, i.e. a larger drag value is more negative than a smaller one), while the term related to lift is here to penalize
asymmetric wakes that could be obtained from consistently biased blowing by the jets in one direction. Such consistently biased blowing strategies can be found when no
lift penalization is used (see Appendix B), but are not desired as they generate large lift values in addition to the reduced drag. In the figures, we
present the value of the drag coefficient, which is a normalized value of the drag $C_D = \frac{-D}{\rho \bar{U}^2 R}$,
where $\bar{U}=2 U(0) / 3$ is the mean velocity magnitude, $\rho$ the volumetric mass density of the fluid, and $R$ the
diameter of the cylinder.

\subsection{Artificial Neural Network and Deep Reinforcement Learning algorithm \label{PPO}}

Artificial intelligence and particularly machine learning have become very
attractive fields of research following several recent high-profile
successes of Deep Artificial Neural Networks (DANNs), such as attaining
super-human performance at image labeling \citep{lecun2015deep}, crushing
human professionals at the game of Go \citep{silver2017mastering}, or
achieving control of complex robots \citep{7989385}, which have shed light
on their ability to handle complex, non-linear systems. Those new techniques
are now being applied to other disciplines, such as Fluid Mechanics
\citep{kutz_2017, brunton2019machine}, and novel applications of DANNs have
recently been proposed for both analyzing laboratory data
\citep{rabault2017performing}, formulation of reduced order models
\citep{srinivasan2019predictions}, AFC
\citep{rabault_kuchta_jensen_reglade_cerardi_2019}, and the control of
stochastic systems from only partial observations \citep{bucci2019control}. In
particular, the Deep Reinforcement Learning (DRL) approach is a promising
avenue for the control of complex systems, including AFC. This approach
leverages DANNs to optimize interaction with the system it should control through
three channels: observing the state of the system, acting to control the system,
and a reward function giving feedback on its current performance. The choice
of the reward function allows to direct the efforts of the DANN towards solving
a specific problem. In the following, we will use the word ``action'' to
describe the value provided by the ANN at each time step based on a state input,
while ``control'' describes the value effectively used in the simulation.

In the present case, similarly to
\citet{rabault_kuchta_jensen_reglade_cerardi_2019}, we use a DANN that is a
simple fully-connected network, featuring one input layer, two consecutive
hidden layers of size $512$ each, and one output layer
providing the action. The classic rectified linear unit (ReLU) is used as an
activation function. The input layer is connected to
$151$ probes immersed in the simulation that measure the value of the pressure
in the vicinity of the cylinder. The output layer provides the action used to
obtain the control, i.e. the mass flow rates of the jets. The reward function,
presented in Eq. \ref{reward_function}, weights both drag and lift to guide the
network towards drag-reducing strategies. The DRL algorithm used for training,
known in the literature as the Proximal Policy Optimization method (PPO,
\cite{schulman2017proximal}), is among the state-of-the-art for training
DANNs to perform continuous control. Each training episode (an episode is a sequence
of interations between the DANN and an independent simulation, generating data to be fed to the PPO algorithm) is
started from a converged, well-defined {K}{\'a}rm{\'a}n vortex street. In the
following an episode has a duration corresponding to around $8$ vortex shedding
periods.

\begin{figure*}[ht]
\begin{center}
\includegraphics[width=0.45\textwidth]{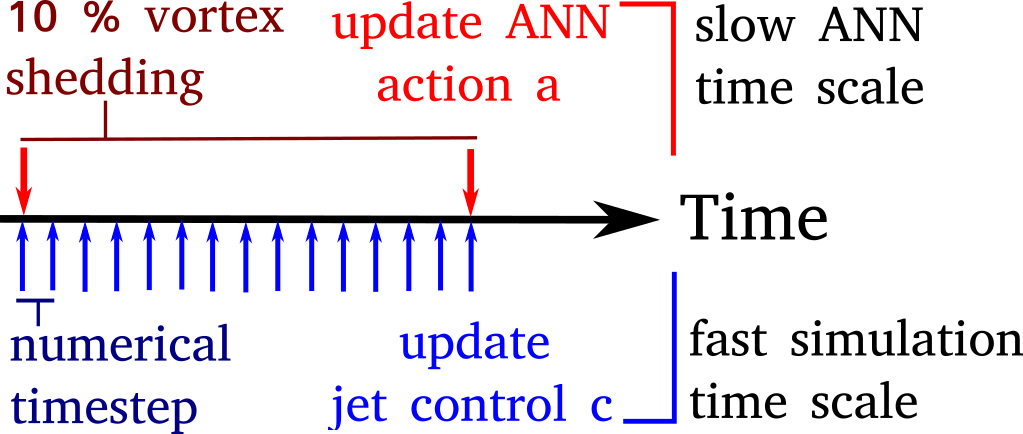}
\caption{Illustration of the existence of two different time scales in the algorithm. The first one is a fast time scale, which corresponds
    to the timestep $dt$ of the simulation, and is imposed by considerations
    around the numerical stability of the simulation. The control applied
    through the jets (i.e. $c$) is updated at this time scale to enforce
    continuity of the quantities used in the solver. The second time scale is a
    slower time scale, which corresponds to around 10\% of the vortex shedding
    period, and captures the relevant time scale of the system. The action set
    by the ANN is updated following this slower
    time scale.}
\label{fig:timescales}
\end{center}
\end{figure*}

It is difficult for the PPO algorithm to learn the necessity to set
time-correlated, continuous actions. This is a consequence of the PPO trying at
first purely random actions. Therefore, we added two limitations to the control
effectively applied in the CFD simulations. First, the action provided by the
network is updated only around typically 10 times per vortex shedding period,
i.e. the action is in practise kept constant for a duration of around typically
$10$\% of the vortex shedding period, except if specifically stated otherwise.
This allows the ANN to update its action at a frequency that corresponds
roughly to the phenomenon to control, rather than at either a frequency too low
(in which case the ANN cannot manage to perform any control), or too high
(in which case the randomness from exploration noise
does not lead to a sufficient net effect on the system, which can reduce the
learning speed). This is a well-known necessity in the literature, and is
analogous to the `frame skip' used in the control of Atari games, for example
\citep{braylan2015frame, neitz2018adaptive}. Second, the control effectively
set in the simulation is obtained from the latest action and previous control
at each numerical time step so that it is made continuous in time to avoid invalid physical
phenomena such as infinite acceleration of the fluid contained in the jets.
To this end, the control at each time step in the simulation is
obtained for each jet as:

\begin{equation}
c_{s+1} = c_{s} + \alpha (a - c_{s}) \label{eqn:set_control},
\end{equation}

\noindent where $c_s$ is the
control of the jet considered at the previous numerical time step, $c_{s+1}$ is the new control, $a$ is the action
provided by the PPO agent for the current set of time steps, and $\alpha = 0.1$ is a numerical parameter. In the present baseline case,
there are 50 updates of the control between two consecutive updates of an action, so the value of $\alpha$ allows a quick convergence
of the control to the action, relative to the time between action updates. In practice,
the exact value of $\alpha$ has little to say for the performance of the control. The existence of those two time scales,
and the update of action versus control, are illustrated in Fig. \ref{fig:timescales}. We provide an illustration of
the effect of both the frequency of action updates and the value of $\alpha$ in Appendix C.

\subsection{Parallelization of the data collection for the DRL algorithm \label{parallel}}

The code released so far (\citet{rabault_kuchta_jensen_reglade_cerardi_2019},
\url{https://github.com/jerabaul29/Cylinder2DFlowControlDRL}) allows to train the ANN to perform AFC in about 24 hours
using a modern CPU running on a single core. The time needed to perform training is one of the weaknesses of this study,
as one would expect that more realistic (and complex) flow simulations may require significantly more learning before
an appropriate control strategy is found. Therefore, obtaining speedups is critical for applying this approach to more
challenging problems. Benchmarking the code revealed that typically about 99.7\% of the computation time is
spent on the environment itself (i.e., the CFD part), rather than the ANN and DRL algorithm. This is due to, on the one hand, the
very efficient ANN and DRL implementations provided by TensorFlow / Tensorforce and, on the other hand, the inherent cost associated with
CFD. In addition, we note that our hardware uses purely CPUs, and that GPU acceleration may even increase this
proportion. Therefore, the optimization effort should focus on the environment part of the code, with Amdahl's Law
predicting that full parallelization would allow a theoretical speedup factor of up to typically 300 for this specific
case.

There are two different ways of obtaining such speedups: first, one can increase the speed of the CFD simulation itself,
i.e. parallelizing the simulation. This topic is well discussed in the literature for a wide range of numerical
methods (for a short introduction on the topic, one may for example refer to \citet{simon1992parallel, GROPP1990289,
GROPP2001337}). However, this approach has its limitations. For example, in our simple 2D simulation, the Finite Element
Model (FEM) problem is small enough that our attempts to parallelize the code lead to very limited speedups (no more
than a factor of typically less than 2, independently of using more CPUs), due to the large amount of communication
needed between the physical cores compared to the small size of the problems to be solved by the individual cores.
This is a well known problem in parallelizing numerical models \citep{GROPP2001337}. Therefore, this option is not
feasible in our case, and even in the case of more complex CFD simulations it will reach a limit, given enough CPUs are available.

Second, one can attempt to parallelize not the DRL / ANN by themselves (as those use only 0.3\% of the computing
time), but the collection of data used to train those algorithms. As the environment
itself cannot be sped up much further, another approach is therefore to use
several environments, with each being an independent, self-sufficient simulation which feeds data in
parallel to the DRL algorithm. This means in practice that the DRL agent learns in parallel from several interactions with
simulations. This results in a very simple parallelization technique, which is well-adapted to cases when most of
the computational time is spent in the environment itself.

More concretely, the PPO algorithm optimizes the expected return of its stochastic policy, which is in practice estimated based on a number $T_L$ of `rollouts', that is, independent episodes of environment interactions which can
be simulated simultaneously. While there exist more sophisticated distributed execution schemes, we observe the following:
if the simulation is by far the dominant bottleneck of the training process (which is usually the case in fluid
mechanics),
the simplest approach is to avoid distributing the (already complex) DRL logic and instead add a lightweight network
communication layer on top of the agent-environment interaction. The DRL framework Tensorforce \citep{kuhnle2017tensorforce}
supports the capability for DRL models to keep track of parallel streams of experience, which in combination with our
environment wrapper allows to parallelize any DRL algorithm for any simulation. In practice, the communication layer
is implemented in Tensorforce and associated with a thin wrapper class that communicates through sockets with the
different environments. This allows to distribute the environments (i.e., in our case, the expensive CFD simulations),
whether on one machine or on a group of machines available through a network if this may be relevant. Those features
are all part of the open-source code release (see Appendix A).

We want to emphasize that our main contribution does not consist of a sophisticated parallelization method for DRL training,
as indeed even the original PPO paper \citep{schulman2017proximal} mentions the parallelized collection of experiences.
Instead we observe that recent approaches to distributing DRL \citep{Horgan2018apex,Espeholt2018impala} are concerned
with massive-scale parallelization, based on the HOGWILD! method \citep{hogwild}. These approaches are useful if the aim is to run 100s of environments simultaneously and if the overhead of message-passing and syncing between different instances, or local cluster of instances, is the bottleneck. Our paper points out that a more straightforward parallelization approach is better suited to the moderately short learning processes but expensive simulations of fluid mechanics problems, which otherwise result in impractical runtimes.
% are mostly `too sophisticated', in that they are concerned with improving aspects irrelevant for the characteristics of fluid mechanics problems.

\section{Results and discussion}

\begin{figure*}[ht]
\begin{center}
\subfloat[]{\includegraphics[width=0.45\textwidth]{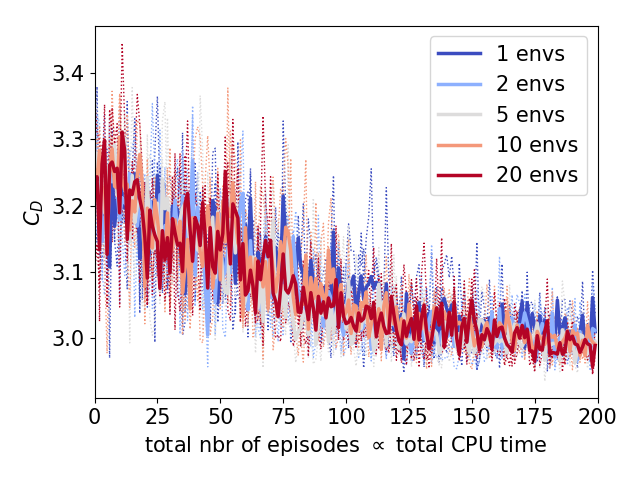}}
\subfloat[]{\includegraphics[width=0.45\textwidth]{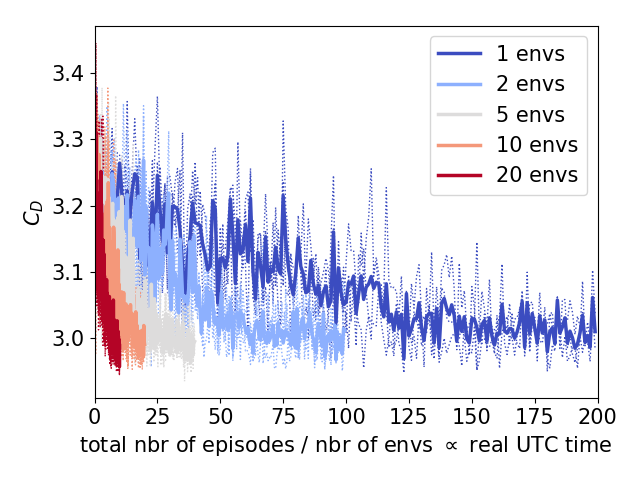}}
\caption{Scaling results obtained by using a total number of environments that is a divider of the update period $T_L$ of the network, in a synced fashion. 3 repetitions are performed for each number of environments.
Individual learnings are indicated by thin lines. The average of all 3 learnings in each case is indicated by a thick
line. In the present case, the learning is formally identical to the serial case (i.e. with one single environment). This is illustrated by (a), which shows that the shapes of the
learning curves are identical
independently of the number of environments used. This naturally results in a perfect speedup, as illustrated by both (a) and (b).}
\label{fig:under}
\end{center}
\end{figure*}

\begin{figure*}[ht]
\begin{center}
\subfloat[]{\includegraphics[width=0.45\textwidth]{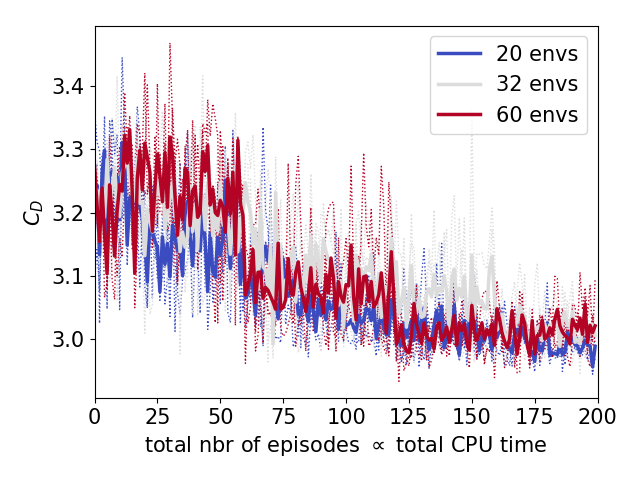}}
\subfloat[]{\includegraphics[width=0.45\textwidth]{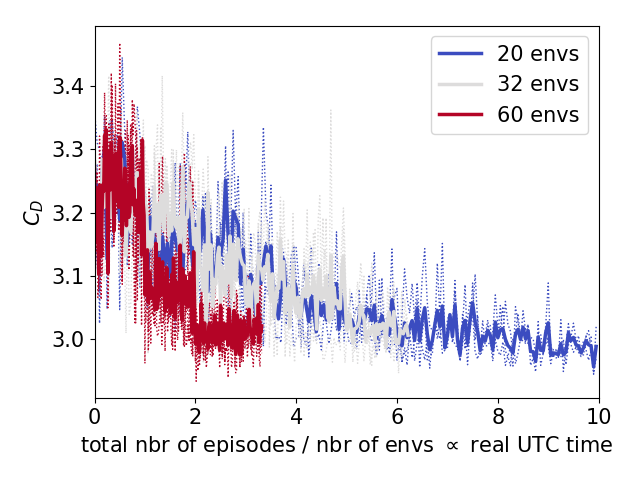}}
\caption{Scaling results obtained by using a total number of environments that is larger than the update period $T_L$. 3
repetitions are performed for each number of environments. Individual learnings are indicated by thin lines. The
average
of all 3 learnings in each case is indicated by a thick line. As visible in (a), learning still takes place almost
satisfactory though steps are clearly visible in the learning curves. This is due to many updates of the network taking
place almost simultaneously, when multiple batches of environments reach the end of an episode at about the same time, which consequently triggers a comparatively large cumulative improvement of the policy. While
this slightly degrades the learning speed measured in terms of raw number of episodes compared with the equivalent synced case of a smaller number of
environments (a), favorable speedups are still observed (b). Note the difference in scale for (b) compared with Fig.
\ref{fig:under} (b), and the final speedup factor obtained (around 55 to 60).}
\label{fig:over}
\end{center}
\end{figure*}

The update period of the network is set to $T_L=20$ episodes, similarly to what was used in
\citet{rabault_kuchta_jensen_reglade_cerardi_2019}. This means that the data
from 20 episodes are gathered between each learning step, as discussed in the previous section. Therefore, using a number of environments that is a divider of
$T_L$, together with the synced running mode (meaning that the DRL algorithm waits for all simulations to be finished
before starting a new series of simulations), will result in a situation that is effectively identical to
the serial learning process. This is illustrated by Fig. \ref{fig:under}, by using respectively 1, 2,
5, 10 and 20 simulation environments. In
Fig. \ref{fig:under}, three runs are performed in each case. The individual performance curves are indicated by thin lines,
while the average of all three runs is indicated by a thicker line. As visible in Fig.
\ref{fig:under} (a), the learning curves using 1, 2, 5, 10 and 20 environments running in parallel all collapse on top of
each other. This
implies that the learning performance is strictly identical between those cases, as we expected. This results in a
perfect speedup by a factor equal to the number of environments, as visible in Fig. \ref{fig:under} (a) and (b).

While perfect scaling is obtained in Fig. \ref{fig:under} as expected from the structure of the algorithm, we are interested in exploring the effect
of further increasing the number of environments on the training quality. This means, in practice, that we ``over-parallelize'' the
collection of data beyond the DRL algorithm's natural parallelization capabilities. Results are presented in Fig. \ref{fig:over},
where we used 32 and 60 environments to investigate the ability of the algorithm to make use of a number of parallel simulations
larger than $T_L$, corresponding to an over-parallelization of by a factor up to 3. This case is not equivalent with serial training, and therefore we do not
enforce synchronization of episodes. As visible in Fig.
\ref{fig:over}, satisfactory speed-up is still observed, although the learning quality is slightly reduced, with the appearance
of clear steps in the learning curves.

These steps are due to the fact that several ANN updates take place in a very
short time interval, when the environments reach the end of an episode. As this happens more or less at
the same time for all environments this means, for example, that three network updates
happen in short succession in case of 60 parallel environments. Overall, this results in a comparatively larger cumulative update of the policy, and that a larger number of
episodes pass in between such updates.

It is interesting to note that the second and third of each of those three consecutive updates are actually
performed based on data that have been collected following an older policy rather than the current policy at time of the second and third update. Therefore, the second and third updates are based on off-policy data.
However, our results indicate empirically that this does not seem to cause problems regarding consistency and stability of the learning
algorithm. This can be explained by the fact that deep learning generally follows an iterative approximate
as opposed to a fully analytic optimization approach. For instance, policy gradient methods like PPO are based on the assumption that a relatively small
number of episode rollouts $T_L=20$ approximate the expected reward well. Moreover, PPO specifically already
performs multiple small updates within one policy optimization step, thus technically using off-policy data for most of these sub-steps.
We conjecture, and observe in our experiments, that such ``slightly off-policy'' updates do not affect the learning process of on-policy DRL algorithms
negatively, which further adds to the effectiveness of our simplistic parallelization approach.

\section{Conclusion}

In this work, we build on the results presented in \citet{rabault_kuchta_jensen_reglade_cerardi_2019} by
providing algorithm parallelization improvements and a multi-environment implementation that greatly speeds up the learning. This allows the DRL algorithm to gather simultaneously data from several independent simulations in a parallel fashion. For a number of environments that is a divider of the
update period of the ANN, the situation is effectively equivalent to classical serial training, and thus perfect scaling is obtained
both theoretically and in practice, up to 20 times in our case. For a number of
environments larger than the update period of the ANN, some of the network updates take place off-policy.
This, together with timing effects in the policy update, causes steps in the learning curve. However, the
PPO algorithm is found to be robust to these ``slightly'' off-policy updates and learning still takes place almost as expected. We experimentally measure
speedups of up to around 60 in this case. We expect there to be a trade-off between extreme over-parallelization and increasingly deteriorating results when comparing to the same effective number of training steps in the serial case.

Those results are an important milestone towards the application of DRL/PPO to Fluid Mechanics problems which are more
sophisticated and realistic than the simple benchmark problem of \citet{rabault_kuchta_jensen_reglade_cerardi_2019}.
Furthermore, the ability to perform training in parallel will allow
to perform several valuable parametric studies in a reasonable amount of time, for example exploring the optimal position of pressure sensors
or how many inputs are necessary for effective control. 
The
scalings observed here are expected to enable even larger performance increases on more complex problems. Indeed, one
usually increases the batch size and the number of episodes between updates together with the underlying complexity of a
DRL task, since more trajectories in the phase space are required to generate meaningful gradient updates with
higher complexity. In addition, the
present parallelization could be combined with the parallelization of the simulation itself, when the size of the
underlying problem is suitable. This is also an important point for getting closer to real-world applications.

To summarize, we foresee that the combination of both parallelism methods could allow DRL of AFC
through CFD to scale to thousands of CPUs, which opens way to applying the methodology to more challenging, and
more realistic, problems. In order to support the development of such methods, all code and
implementations are released as open-source (see Appendix A).

\section{Acknowledgement}

The help of Terje Kvernes for setting up the computational infrastructure used in this work is gratefully acknowledged.
In addition, we want to thank Dr. Miroslav Kuchta for many interesting discussions and an early version of the socket
communication. We gratefully acknowledge help, guidance, and CPU time provided by UH-IaaS (the Norwegian Academic Community
Cloud, IT departments, University of
Oslo and University of Bergen. http://www.uh-iaas.no/). This work was performed thanks to funding received by the University of Oslo in the
context of the 'DOFI' project (grant number 280625).

\section{Appendix A: Open Source code}

The source code of this project, together with a docker container that enforces full reproducibility of our results, is
released as open-source on the GitHub of the first author [NOTE: the repository is empty for now, the code will be released upon
publication in the peer-reviewed literature]:
\textit{https://github.com/jerabaul29/Cylinder2DFlowControlDRLParallel}. The simulation environment is based on the open-source
finite element framework FEniCS \citep{logg2012automated} v 2018.1.0. The PPO agent is based on the open-source
implementation provided by Tensorforce \citep{kuhnle2017tensorforce}, which builds on top of the Tensorflow
framework \citep{abadi2016tensorflow}. This code is an extension of the serial
DRL for active flow control available here: \textit{https://github.com/jerabaul29/Cylinder2DFlowControlDRL}.

\section{Appendix B: learning with and without lift penalization}

\begin{figure*}[ht]
\begin{center}
\includegraphics[width=0.65\textwidth]{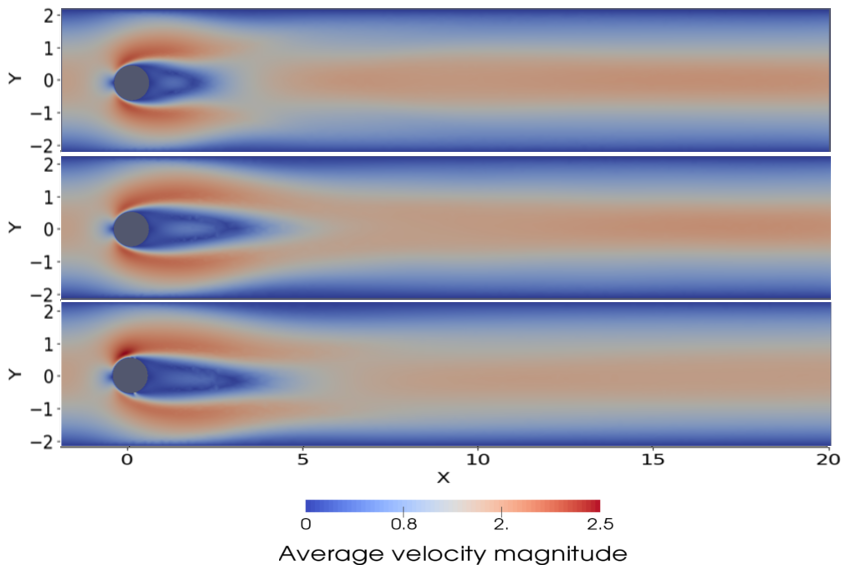}
\caption{Illustration of the effect of 'cheating' strategies when no lift penalization is present. All velocity maps indicate
    the velocity magnitude, averaged over around 10 vortex shedding periods. At the top, no control is present. In the middle,
    control is present, and the DRL algorithm was trained with lift penalization. In this case, control is taking place satisfactorily
    and the jets are used to control the vortex shedding. At the bottom, control is present and no lift penalization is used during training (in addition,
    the maximum intensity of the jets is allowed to be larger so that the effect of biased blowing is more visible). In this case, clear asymmetry of the
    wake and velocity pattern is visible, and the velocity excess present at the top of the cylinder compared to the bottom creates a
    large mean lift value.}
\label{fig:illustration_cheating}
\end{center}
\end{figure*}

In the case when no lift penalization is used in the equation for the reward Eqn. (\ref{reward_function}), the DRL
algorithm is able to increase its reward through discovering strategies in which a systematic bias is present, i.e.
jets blow consistently in the same direction at their maximum strength after a given point in time. This is considered as a 'cheating' strategy,
as the drag reduction is accompanied by the production of a large lift. This problem is illustrated in Fig. \ref{fig:illustration_cheating}.
This problem is effectively suppressed by applying the lift penalization presented in Eqn. (\ref{reward_function}), since
this applies a negative reward to large lift biases introduced by this kind of strategies.

\section{Appendix C: effect of action frequency and control smoothing}

\subsection{Effect of the action update frequency}

\begin{figure*}[ht]
\begin{center}
\includegraphics[width=0.45\textwidth]{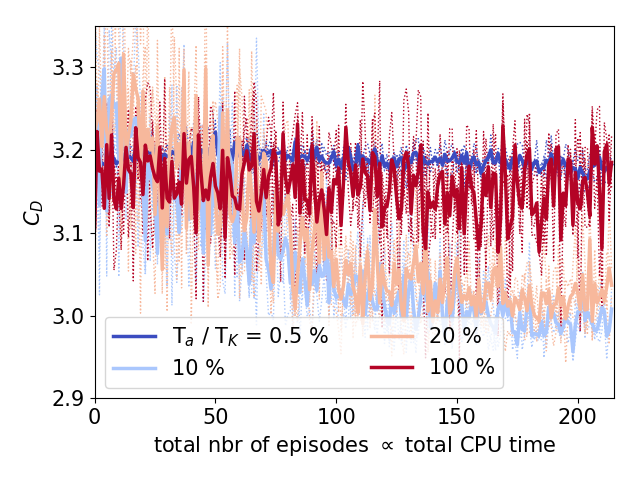}
    \caption{Illustration of the importance of the choice of a relevant time
    scale for the update of the action set by the network (action update is
    performed with a period $T_a$), relatively to the time scale describing the
    dynamics of the system to control (the Karman alley has a typical time
    scale described by the period $T_K$). A too low $T_a$ means that the random
    exploration does not create any consistent forcing that changes the state
    of the system, and therefore no learning takes place. A too high $T_a$
    means that the update in the action is too slow to perform a control
    varying with the phase of the vortex shedding, and therefore no control can
    take place. In the middle, there is a sweet spot where control is possible
    and learning takes place. As in the other similar figures, individual
    learnings are indicated by thin lines. The average of all 3 learnings in
    each case is indicated by a thick line.}
\label{fig:showing_different_time_scales}
\end{center}
\end{figure*}

As explained in Fig. \ref{fig:timescales} and section II B, there are two
different natural time scales: the first one is very short and corresponds to
the numerics of the CFD simulation (typical period associated with the numerics
of the simulation: $dt$), and the second one is slower and corresponds to the
physical dynamics happening in the system (typical period of the Karman vortex
shedding: $T_K$). Therefore, the control applied to the simulation must be
applied at an intermediate time scale $T_a$ such that $T_K > T_a > dt$. A bad
choice of $T_a$ makes learning impossible, as illustrated in Fig.
\ref{fig:showing_different_time_scales}. In the case when $T_a / T_K = 0.5$\%,
the action frequency is too high, and therefore the random exploration noise is
applied for a very short time before being updated. This means that the ANN is
never able to significantly modify the state of the Karman vortex street and
cannot learn. By contrast, in the case when $T_a / T_K = 100$\%, the action
update frequency is too low for the ANN to control anything of the vortex
dynamics. In the middle of those values, there is a sweet spot at around
$T_a/T_K=10$\% where the frequency of action update is high enough so that the
ANN can control the system, but low enough so that the random exploration noise
has enough time to act before update so that the system reacts to it in a
measurable way.

\subsection{Effect of the smoothing law}

\begin{figure*}[ht]
\begin{center}
\includegraphics[width=0.45\textwidth]{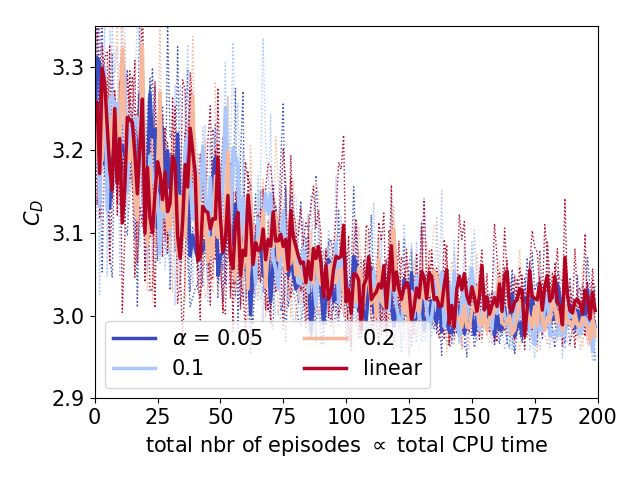}
    \caption{Illustration of the stability of the learning process respectively to the
    exact value of $\alpha$ and, more generally, the kind of interpolation used
    for computing the control between action updates by the ANN. As in the
    other similar figures, individual learnings are indicated by thin lines.
    The average of all 3 learnings in each case is indicated by a thick line.
    }
\label{fig:showing_effect_smoothing}
\end{center}
\end{figure*}

As a consequence of the necessity to choose a relevant action time scale $T_a$
that is much larger than the numerical simulation time scale $dt$, one needs
to interpolate between action updates to generate the control to use at each
time step. This is schematically shown by Fig. \ref{fig:timescales}, and also explained in
section II B. The interpolation can be performed in several fashions, and must
follow some conditions of smoothness and continuity to avoid both unphysical
phenomena such as infinite accelerations in the fluid, and phenomena of
numerical instability. One can for example use an exponential decay law based
on the control value from the previous action to the new one, corresponding for
example to what is used in Eqn. (\ref{eqn:set_control}). In this case, which is
the one used in this whole paper unless explicitly stated differently, the
choice of the decay constant allows to adapt the speed of the convergence of
the control towards the value given by the new action. In our case, we find
that reasonable values lead to satisfactory learning (see Fig. \ref{fig:showing_effect_smoothing}). Other
laws can be used, for example linear interpolation between the actions
determining the controls. This also works fine, as illustrated in Fig. \ref{fig:showing_effect_smoothing}.
A careful observation of Fig. \ref{fig:showing_effect_smoothing} can suggest that
the linear interpolation is slightly less efficient than the exponential decay, and
this may be a consequence of the fact that the linear interpolation converges to
the value of the updated action slower than the exponential decay, which may introduce
a form for lag in the control by the ANN. However, this phenomenon is quite minor.

\bibliographystyle{jfm}
\bibliography{template}

\end{document}